\documentclass[10pt,aps,prl,twocolumn,superscriptaddress,postprint]{revtex4-2}

\usepackage[english]{babel}
\usepackage{physics}
\usepackage{graphicx}  
\usepackage{amssymb}
\usepackage{amsmath}  
\usepackage{dsfont}
\usepackage{float}
\usepackage{tabularx}
\usepackage{multirow}
\usepackage{array}
\usepackage{booktabs}
\usepackage{stmaryrd}
\usepackage{mathrsfs}
\usepackage{tikz}
\usepackage{mathtools}
\usepackage{hyperref}
\usepackage{cleveref}
\usepackage{orcidlink}
\usepackage{bm}
\usepackage{csquotes}
\usepackage{siunitx}

\addto\captionsenglish{%
}
\addto\captionsenglish{%
}

\newcommand{\figref}[2][{}]{Fig.~\hyperref[#2]{\ref{#2}#1}}
\newcommand{\tabref}[2][{}]{\tablename~\hyperref[#2]{\ref{#2}#1}}
\hypersetup{
  colorlinks = true,
  citecolor = blue,
  linkcolor = blue,
  urlcolor = black,
}

\usetikzlibrary{backgrounds, positioning, arrows.meta, calc}

\DeclarePairedDelimiter{\superbra}{\langle\!\langle}{\rvert}
\DeclarePairedDelimiter{\superket}{\lvert}{\rangle\!\rangle}
\DeclarePairedDelimiterX{\superskp}[2]{\langle\!\langle}{\rangle\!\rangle}{#1 \delimsize\vert #2}

\begin{document}

% % Title, Author, Date
\title{Bridging Quantum Noise and Classical Electrodynamics with Stochastic Methods}

\author{Felix \surname{Hitzelhammer}\,\orcidlink{0009-0006-5335-7363}}
\affiliation{Institute of Physics, NAWI Graz, \href{https://ror.org/01faaaf77}{\textcolor{blue}{University of Graz}}, 8010~Graz, Austria}
\email[Contact author:~]{felix.hitzelhammer@uni-graz.at}
\author{Johannes~\surname{Stowasser}\,\orcidlink{0009-0007-2240-163X}}
\affiliation{TUM School of Computation, Information and Technology, \href{https://ror.org/02kkvpp62}{\textcolor{blue}{Technical University of Munich}}, 85748 Garching, Germany\looseness=-1}
\author{Lukas~\surname{Hanschke}\,\orcidlink{0000-0002-1943-7603}}
\affiliation{TUM School of Computation, Information and Technology, \href{https://ror.org/02kkvpp62}{\textcolor{blue}{Technical University of Munich}}, 85748 Garching, Germany\looseness=-1}
\affiliation{Walter Schottky Institut, \href{https://ror.org/02kkvpp62}{\textcolor{blue}{Technical University of Munich}}, 85748 Garching, Germany\looseness=-1}
\author{Katarina~\surname{Boos}\,\orcidlink{0000-0003-3560-9134}}
\affiliation{TUM School of Computation, Information and Technology, \href{https://ror.org/02kkvpp62}{\textcolor{blue}{Technical University of Munich}}, 85748 Garching, Germany\looseness=-1}
\affiliation{Walter Schottky Institut, \href{https://ror.org/02kkvpp62}{\textcolor{blue}{Technical University of Munich}}, 85748 Garching, Germany\looseness=-1}
\author{Tobias~C.~\surname{Sutter}\,\orcidlink{0009-0008-3179-7979}}
\affiliation{Faculty of Physics, \href{https://ror.org/03prydq77}{\textcolor{blue}{University of Vienna}}, 1090~Vienna, Austria}
\author{Michael~\surname{Haider}\,\orcidlink{0000-0002-5164-432X}}
\affiliation{TUM School of Computation, Information and Technology, \href{https://ror.org/02kkvpp62}{\textcolor{blue}{Technical University of Munich}}, 85748 Garching, Germany\looseness=-1}
\author{Christian~\surname{Jirauschek}\,\orcidlink{0000-0003-0785-5530}}
\affiliation{TUM School of Computation, Information and Technology, \href{https://ror.org/02kkvpp62}{\textcolor{blue}{Technical University of Munich}}, 85748 Garching, Germany\looseness=-1}
\affiliation{TUM Center for Quantum Engineering (ZQE), 85748 Garching, Germany}
\author{Kai~\surname{Müller}\,\orcidlink{0000-0002-4668-428X}}
\affiliation{TUM School of Computation, Information and Technology, \href{https://ror.org/02kkvpp62}{\textcolor{blue}{Technical University of Munich}}, 85748 Garching, Germany\looseness=-1}
\affiliation{Walter Schottky Institut, \href{https://ror.org/02kkvpp62}{\textcolor{blue}{Technical University of Munich}}, 85748 Garching, Germany\looseness=-1}
\affiliation{TUM Center for Quantum Engineering (ZQE), 85748 Garching, Germany}
\affiliation{Munich Center for Quantum Science and Technology (MCQST), 80799 Munich, Germany}
\author{Gabriela~\surname{Slavcheva}\,\orcidlink{0000-0001-5474-9808}}
\affiliation{Institute of Physics, NAWI Graz, \href{https://ror.org/01faaaf77}{\textcolor{blue}{University of Graz}}, 8010~Graz, Austria}
\affiliation{Quantopticon, Chicago, Illinois 60615, USA}
\author{Ulrich~\surname{Hohenester}\,\orcidlink{0000-0001-8929-2086}}
\affiliation{Institute of Physics, NAWI Graz, \href{https://ror.org/01faaaf77}{\textcolor{blue}{University of Graz}}, 8010~Graz, Austria}

\date{\today}

\begin{abstract}
The development of emerging technologies in quantum optics demands accurate models that faithfully capture genuine quantum effects. Mature semiclassical approaches reach their limits when confronted with quantized electromagnetic fields, while full Hilbert space treatments are often computationally prohibitive. To address these challenges, we develop a framework based on coupled stochastic processes with a common cross-covariance structure that can be easily coupled to various types of Maxwell solvers. Our approach accounts for the non-commutativity in the quantum-to-classical transition in a natural way, and has the ability to capture quantum optical signatures while retaining compatibility with classical electromagnetics. For benchmarking, we compare our simulation results with experimental emission spectra of a strongly driven InGaAs quantum dot, finding excellent agreement. Our results highlight the potential of tailored stochastic processes for simulating non-classical light in complex photonic environments.
\end{abstract}

\maketitle

\section{Introduction}\label{sec:introduction} 
Modeling light-matter interactions on the nanoscale across different architectures in quantum photonics is essential for advancing the design and fabrication of the key components of many emerging quantum technologies~\cite{wang2025scalable,damico2019nanoscale,flamini2018photonic,lodahl2015interfacing,heindel2023quantum}. Prominent examples include quantum communications, photonic quantum computing, and high-precision quantum sensing~\cite{gisin2002quantum,wang2020integrated,obrien2007optical,giovannetti2004quantum-enhanced}. Classical electromagnetic solvers based on Maxwell’s equations provide a powerful framework for simulating complex nanostructures~\cite{joannopoulos2008photonic,burger2012finite,taflove2005computational}, but they are inherently restricted to classical light propagation and cannot capture nonclassical features such as photon correlations or entanglement. These phenomena require quantum simulation approaches, typically formulated through (quantum) master equations or related quantum-optical models implemented on classical computers. However, these models often describe idealized systems and cannot account for propagation and quantum interference effects in realistic extended macroscopic devices and circuits with different materials, geometries and boundary conditions. A unified approach that bridges quantum and classical descriptions is therefore highly desirable and of long-standing foundational interest~\cite{wigner1932quantum,carmichael1999statistical,rundle2021overview}.

On the one hand, exact quantum descriptions, such as the Gorini-Kossakowski-Sudarshan-Lindblad formalism~\cite{lindblad1976generators,gorini1976completely} or quantum stochastic differential equations~\cite{hudson1984quantum}, are both elegant and fundamental, but they do not provide a natural route to couple to classical Maxwell solvers and rapidly become computationally prohibitive for large or complex systems due to exponential Hilbert-space growth~\cite{schollwoeck2011density-matrix,breuer2002theory}. Furthermore, semiclassical approximations that allow coupling to electromagnetic solvers, e.g., based on the Maxwell-Bloch equations \cite{allen1987optical,jirauschek2019optoelectronic} or more advanced Maxwell-Bloch Langevin approaches~\cite{popp2024modeling,pusch2012coherent,andreasen2010numerical,slavcheva2004fdtd}, must be used with caution, as they cannot accurately capture quantum features such as non-classical photon statistics and correlations. This motivates the use of stochastic differential equations (SDEs), which are intrinsically driven by noise, to capture correlations, fluctuations, and noise spectra beyond average dynamics.

The derivation of suitable SDEs from the underlying quantum dynamics employs phase-space representations \cite{lee1995theory,sudarshan1963equivalence,husimi1940formal}, quantum trajectory methods \cite{plenio1998quantum-jump,molmer1993monte,carmichael1993quantum}, or quantum Langevin equations \cite{gardiner2004quantum,louisell1973quantum}. Phase-space methods, such as the Wigner or positive $P$ representations, map the density matrix onto a quasi-probability distribution, which is governed by a Fokker-Planck equation (FPE). By extracting drift and diffusion from this FPE, one obtains the corresponding SDE~\cite{risken1966quantum,scully1997quantum,haken1984laser,chang1985quantum}. Quantum trajectory methods, also known as Monte Carlo wave-function approaches, describe the system in terms of individual stochastic realizations of its wavefunction. They naturally incorporate the randomness associated with quantum jumps, continuous monitoring, and measurement back-action~\cite{plenio1998quantum-jump,molmer1993monte,carmichael1993quantum}. Quantum Langevin equations offer an operator-level description of system-bath interactions. Under suitable approximations, such as linearization or semiclassical limits, they can be recast as stochastic differential equations for classical variables (c-numbers)~\cite{gardiner2004quantum,louisell1973quantum,carmichael1999statistical,carmichael2008statistical}. Here, we introduce an SDE approach, without relying on wave-function unraveling, phase-space representations, or operator-valued noise.

As a first validation of our approach, we consider the Mollow spectrum, which serves as a compelling example of the successful interplay between theory and experiment~\cite{mollow1969power,schuda1974observation}. It arises from the resonance fluorescence of a strongly driven two-level system (TLS)~\cite{meystre2007elements}, exhibiting a characteristic triplet structure that reflects coherent light-matter interactions. Excitonic transitions in solid-state systems, such as semiconductor quantum dots (QDs), can be modeled as effective TLSs, and the associated Mollow spectrum has been experimentally observed~\cite{flagg2009resonantly,vamivakas2009spin-resolved}. The Mollow spectrum not only benchmarks theoretical models but also illustrates the importance of capturing fluctuations and correlations in open quantum systems~\cite{chen2019predective}. Beyond providing a benchmark for theoretical models, the sidebands in resonance fluorescence can exhibit non-classical photon correlations, highlighting the relevance of the Mollow triplet for quantum-optical control and photon-correlation experiments~\cite{lopez2017photon,steck2007quantum}.
The Mollow triplet emission provides a promising approach for single-photon and sequential photon generation \cite{ulhaq2012cascaded}, as well as for light-matter entangled-state generation \cite{carreno2024entanglement,vamivakas2009spin-resolved}.

Motivated by the incompatibility between quantum simulations and Maxwell solvers, we introduce a stochastic Bloch-vector approach underpinned by SDEs, and validate it by comparison to dedicated experiments. This method links the system’s quantum dynamics to electromagnetic fields, providing a bridge between quantum emitter dynamics and spatio-temporally resolved classical fields. We show that SDEs can properly account for the dynamics of open quantum systems and offer the potential for integration with Maxwell solvers~\cite{stowasser2024stochastic,popp2024modeling,park2024stochastic}. Finally, we argue that deriving a stochastic Bloch vector from a non-symmetric diffusion matrix addresses a critical gap in the stochastic modeling of systems subject to non-trivial noise processes.

%%%%%%%%%%%%%%%%%%%%%%%%%%%%%%%%%%%%%%%%%%%%%%%%%%%%%%%%%%%%%%%%%%%%%%%%%%%%%%%%%%%%%%%%%%%%%%%%%%%%%%%%%%%%%%%%%%%%%%%%%%%%%%%%%%%%RESULTS%%%%%%%%%%%%%%%%%%%%%%%%%%%%%%%%%%%%%%%%%%%%%%%%%%%%%%%%%%%%%%%%%%%%%%%%%%%%%%%%%%%%%%%%%%%%%%%%%%%%%%%%%%%%%%%%%%%%%%%%%%%%%%%%%%%%%%%%%%%%%%%%%%%%%%%%%%%%%%%%%%%%

\begin{figure}
    \centering
    \includegraphics[width=1\linewidth]{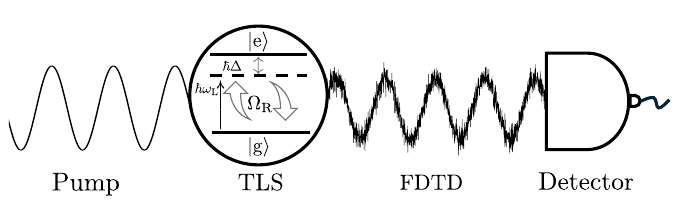}
    \caption{Schematic representation of the modeled system. From left to right: The two-level system (TLS), embedded in vacuum, is coherently driven by a classical field, inducing Rabi oscillations at the Rabi frequency $\Omega_{\mathrm{R}}$ between the ground state $\ket{g}$ and excited state $\ket{e}$. The detuning energy $\hbar\Delta$ quantifies how far the driving frequency is off-resonant from the natural transition frequency of the TLS, and $\hbar\omega_{\mathrm{L}}$ denotes the laser photon energy. The stochastic trajectories of the TLS steady-state dynamics are coupled to a source current and are propagated using a finite-difference time-domain (FDTD) solver.}
    \label{fig:schmeatic_representation}
\end{figure}

\begin{figure*}
    \centering
    \hspace*{-0.8cm}
    \includegraphics[scale=1.05]{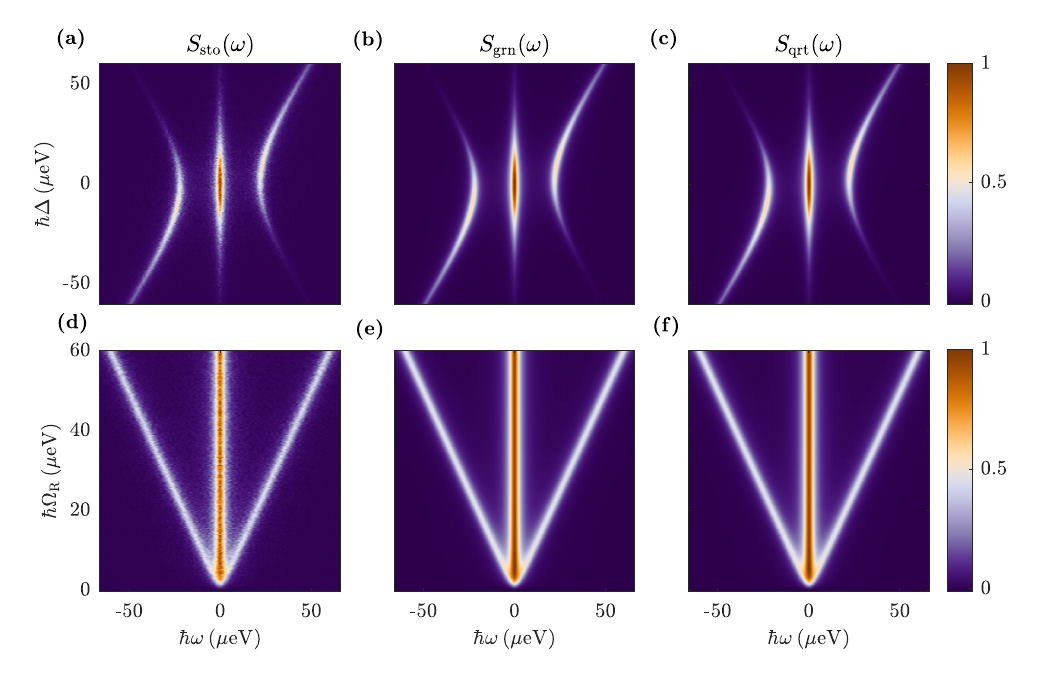}
    \caption{Maps of Mollow spectra obtained (from left to right) by stochastic approach, Green's propagator and quantum regression theorem. The presented data is normalized to their respective maxima. Panels (a--c) show the Mollow spectrum $S(\omega)$  in dependence of the detuning energy $\hbar \Delta$, and (d--f) as a function of the Rabi energy $\hbar \Omega_{\mathrm{R}}$. Parameters used in the simulations are listed in Table~\ref{Simulation parameters}.}
    \label{fig:theoretical_Mollow_spectra_colormap}
\end{figure*}

\begin{figure*} 
    \centering
    \hspace*{-0.8cm}
    \includegraphics[scale=1]{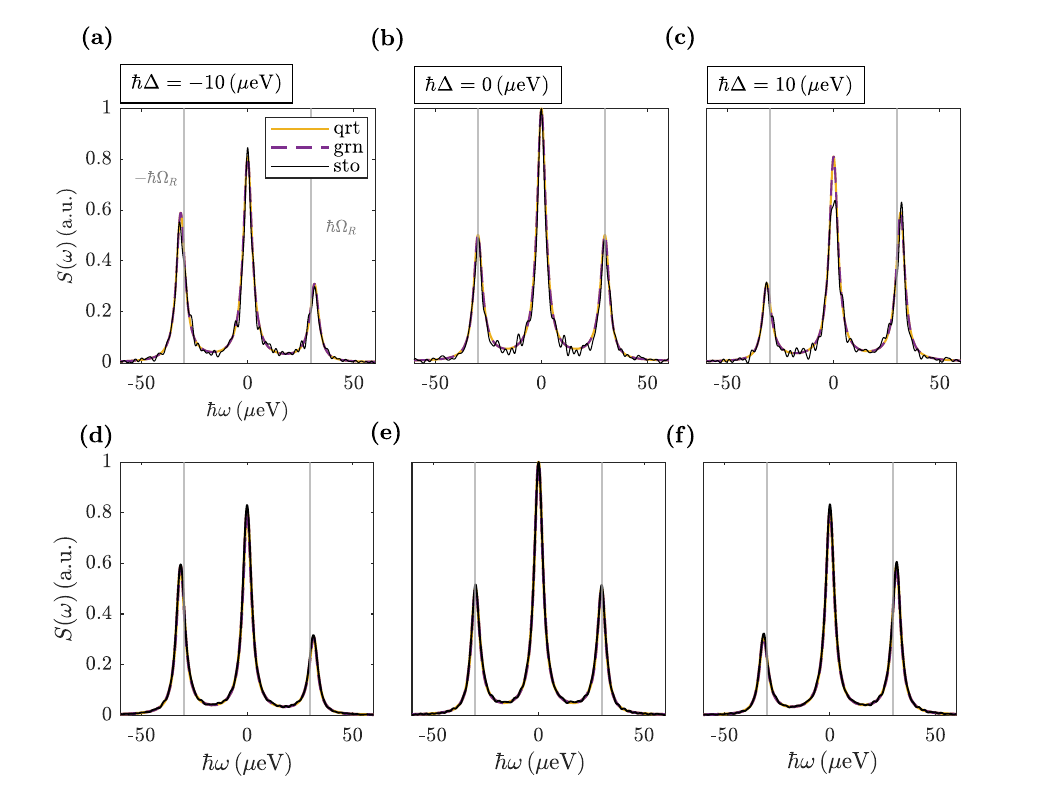}
    \caption{Lineplots of Mollow spectra obtained by quantum regression theorem (yellow line), Green's propagator (purple dashed line) and stochastic approach (black line). The presented data is normalized to their respective maxima. The grey lines indicate the Rabi energy $\hbar\Omega_{\mathrm{R}}$. Panels (a-f) show the Mollow spectrum $S(\omega)$ at a fixed Rabi energy with $\hbar\Omega_{\mathrm{R}}$ and varying detuning energy $\hbar\Delta$. The increase of the Walkers $N$ from (a-c) $N = 4000$, to (d-f) $N = 100000$ shows the convergence of the stochastic approach. Simulation parameters: Table~\ref{Simulation parameters}.}
    \label{fig:theoretical_Mollow_spectra_lineplot}
\end{figure*}

\section{Results}\label{sec:results}

\subsection{SDEs for symmetric vs. non-symmetric diffusion matrices}\label{sec:SDEs_for_open_quantum_systems}

We start by briefly summarizing the standard path from a (quantum) master equation to an SDE. After fixing the operator ordering, the master equation can be mapped to a phase-space quasi-probability distribution by replacing quantum operators with corresponding differential operators~\cite{gerry2024introductory,gardiner2004quantum}. From there, one derives an FPE, whose main building blocks are the drift matrix $\bm{A}$, responsible for the deterministic evolution, and the diffusion matrix $\bm{D}$, describing the stochastic fluctuations~\cite{pavliotis2014stochastic}. In conventional treatments, $\bm{D}$ is typically symmetric and positive semidefinite, allowing a factorization~\cite{pavliotis2014stochastic,gardiner2021elements}
\begin{equation}\label{eq:factorization}
    \bm{D} = \bm{B}\bm{B}^{\mathrm{T}}\,,
\end{equation}
that defines the noise coupling for the SDE via the noise matrix $\bm{B}$. This factorization is not unique, a property that has been exploited in previous work~\cite{deuar2002gauge,rousse2023simulations}. The resulting linear Itô SDE with additive noise is~\cite{pavliotis2014stochastic}
\begin{equation}\label{eq:SDE}
    \partial_{t}\bm{s}(t) = \bm{A}\bm{s}(t)+\bm{B}\bm{\xi}(t)\,, 
\end{equation}
where $\bm{s}(t)$ is the stochastic process that solves the SDE and $\bm{\xi}(t)$ is a vector-valued Gaussian white noise process. % This procedure can be summarized schematically in the following way:
%
%\medskip
%
%
%\begin{tikzpicture}[node distance=0.2cm, every node/.style={font=\sffamily}, >=Stealth]
%
% Nodes
%\node (master) {Master eq.};
%\node (phase) [right=of master] {Phase-space};
%\node (fp) [right=of phase] {Fokker-Planck eq.};
%\node (SDE) [right= of fp] {SDE};
%
% Arrows (curved)
%\draw[->, bend left=30] (master) to node[above] {} (phase);
%\draw[->, bend left=30] (phase) to node[above] {} (fp);
%\draw[->, bend left=30] (fp) to node[above] {} (SDE);
%
% Small circular label nodes
%\node[circle, draw, inner sep=1pt, font=\tiny, align=center, above] at ($(master)!0.5!(phase) + (0,0.6)$) {Replace operator\\ with differential\\operator};
%\node[circle, draw, inner sep=1pt, font=\small, above] at ($(phase)!0.5!(fp) + (0,0.6)$) {$\bm{A},\,\bm{D}$};
%\node[circle, draw, inner sep=2pt, font=\scriptsize, align=center, above] at ($(fp)!0.5!(SDE) + (0,0.6)$) $\bm{D}=\bm{B}\bm{B}^{\mathrm{T}}$};
%
%\end{tikzpicture}
%
However, the restriction to a symmetric $\bm{D}$ excludes crucial asymmetric noise correlations originating from non-commuting operators in the quantum realm. In the following, we show how to overcome this limitation by allowing a complex and non-symmetric $\bm{D}$ which can be interpreted as a noise cross-covariance between two coupled stochastic processes.

As a representative example, we consider a TLS driven coherently by a laser field (see Fig.~\ref{fig:schmeatic_representation}). With the help of the pseudospin operators $\sigma_{\pm}$ and $\sigma_{z}$ it can be described by the Hamiltonian~\cite{hohenester2020nano}
\begin{equation}\label{eq:Hamiltonian}
    H = \frac{\hbar}{2} \bigl( \Delta \sigma_{\mathrm{z}} - \Omega_{\mathrm{R}} \left( \sigma_{+} + \sigma_{-} \right) \bigr) \,,
\end{equation}
where $ \Omega_{\mathrm{R}} $ denotes the Rabi frequency, and $ \Delta = \omega_{0} - \omega_{\mathrm{L}} $ is the detuning of the two-level transition frequency $\omega_{0}$ with respect to the laser frequency $ \omega_{\mathrm{L}} $. The TLS dynamics is governed by the standard Lindblad master equation in the Born-Markov and rotating wave approximation~\cite{breuer2002theory,hohenester2020nano}. We choose the Lindblad jump operators $L_{1}$ and $L_{2}$ such that they take spontaneous emission and dephasing into account. 
Moving to Liouville space (see Methods), the Liouvillian superoperator in the Schrödinger picture~\cite{gyamfi2020fundamentals},
\begin{equation}\label{eq:Liouvillian}
    \mathcal{L} = -\mathrm{i}\hbar^{-1} \llbracket H, \mathds{1}_{2} \rrbracket + \sum_{k=1}^{2} \Bigl( L_{k} \otimes L_{k}^{*} - \frac{1}{2}\llbracket  L_{k}^{\dagger}L_{k}, \mathds{1}_{2}\rrbracket_{+} \Bigr) \,,
\end{equation}
acts on the vectorized density matrix $\superket{\rho(t)}$. The steady state $\superket{\rho_{\text{s.s.}}}$ is obtained from the zero-eigenvalue eigenmode of $\mathcal{L}$~\cite{hohenester2020nano,albert2014symmetries}, 
\begin{equation}\label{eq:rho_steady_state}
\superket{\rho_{\mathrm{s.s.}}}
= \lim_{t \to \infty} \sum_{k} \mathrm{e}^{\lambda_{k}t}
\superket{r_{k}}
\superskp{l_{k}}{\rho(0)} \,,
\end{equation}
where $\lambda_k$ are the complex eigenvalues of $\mathcal{L}$ with corresponding biorthonormal right and left eigenvectors $\superket{r_k}$ and $\superbra{l_k}$. Here, the initial state $\superket{\rho(0)}$ is chosen to be the ground-state supervector $\superket{\sigma_{\mathrm{gg}}}$. Based on this, we define a second-order cumulant matrix $\bm{M}= (M_{ij}) \in \mathbb{C}^{3\times3}$ for the pseudospin operators, which captures correlations in the steady state. Its entries are given by
\begin{equation} \label{eq:second_order_cumulant_matrix}
    M_{ij} = \superskp{(\sigma_{i}\sigma_{j})^{\dagger}}{\rho_{\mathrm{s.s.}}} - \superskp{\sigma_{i}^{\dagger}}{\rho_{\mathrm{s.s.}}} \superskp{\sigma_{j}^{\dagger}}{\rho_{\mathrm{s.s.}}} \,,
\end{equation}
where $i,j\in\{-,+,z\}$.

The system’s dynamics can be recast into two coupled linear SDEs, each driven by additive noise with a shared cross-covariance structure. We start by defining a non-symmetric diffusion matrix $\bm{D}$ as a linear combination of the second-order cumulant matrix $\bm{M}$ and the drift matrix $\bm{A}$,
\begin{equation}\label{eq:noise_cross_covariance}
    - \bm{D}=\bm{A}\bm{M}+\bm{M}\bm{A}^{\mathrm{T}} \,,
\end{equation}
which is reminiscent of an algebraic Lyapunov equation~\cite{pavliotis2014stochastic}. This observation fits well into recent work, which has revealed a deep connection between Lyapunov equations and open quantum systems~\cite{purkayastha2022lyapunov}. Since in Eq.~\eqref{eq:noise_cross_covariance}, $\bm{D}$ is a complex and non-symmetric matrix, it loses the meaning of a diffusion matrix as used in SDE theory. Instead, it turns out that $\bm{D}$ acts as the noise cross-covariance matrix. In order to factorize $\bm{D}$, we use the singular value decomposition (SVD) over the commonly used Takagi factorization~\cite{rousse2023simulations}, which enables us to decompose non-symmetric matrices by
\begin{equation}\label{eq:SVD}
    \bm{D} = \bm{U} \, \bm{\Sigma} \, \bm{V}^{\dagger} \, ,
\end{equation}
where $\bm{U}, \bm{V}$ are unitary matrices, and $\bm{\Sigma}$ is a diagonal matrix with non-negative real entries. We introduce the corresponding complex noise matrices 
\begin{equation}\label{eq:noise_matrices}
     \bm{B}_{1} = \bm{U} \, \sqrt{\bm{\Sigma}}\,, \quad   \bm{B}_{2} =  \bm{V}^{*} \sqrt{\bm{\Sigma}}\,, 
\end{equation}
such that $\bm{B}_{1}\bm{B}_{2}^{\mathrm{T}}=\bm{D}$. The noise matrices $\bm{B}_{1}$ and $\bm{B}_{2}$ together with the drift matrix $\bm{A}$ define a \enquote{doubled} system of linear SDEs with inhomogeneity $\bm{b}$ and additive noise (in Itô form, see Methods for details)
\begin{equation} \label{eq:SDE_physics_style}
    \partial_{t}\bm{s}_{k}(t) = \bm{A}\bm{s}_{k}(t) + \bm{b}  + \bm{B}_{k}\,\bm{\xi}(t)\,, \quad k \in\{1,2\}\,,
\end{equation}
where for each $k$ the vector-valued stochastic process $\bm{s}_{k}=(s_{k,-}\,\,s_{k,+}\,\,s_{k,z})^{\mathrm{T}}$, indexed in analogy to pseudospin operators, solves Eq.~\eqref{eq:SDE_physics_style}. Importantly, as we will see in the next section, this is not merely a labeling. Rather, the components of $\bm{s}_{k}$ fulfill an analog modeling role as their operator-valued counterparts. This representation allows us to interpret the dynamics as complex stochastic Bloch vector trajectories, whose ensemble behavior reproduces the correct steady-state correlations. To our knowledge, this use of an SVD factorization to construct two coupled SDEs with a prescribed cross-covariance has not been discussed in the quantum-optics literature. Most importantly, from the ensemble average of the stochastic Bloch vectors $\bm{s}_{1}(\tau),\,\bm{s}_{2}(0)$ we can compute two-time correlation functions, which would not be possible in a purely semiclassical approach. As we will demonstrate next, this allows us to simulate quantum features in optical experiments.
\begin{figure*}
    \centering
    \hspace*{-0.2cm}
    \includegraphics[scale=0.83]{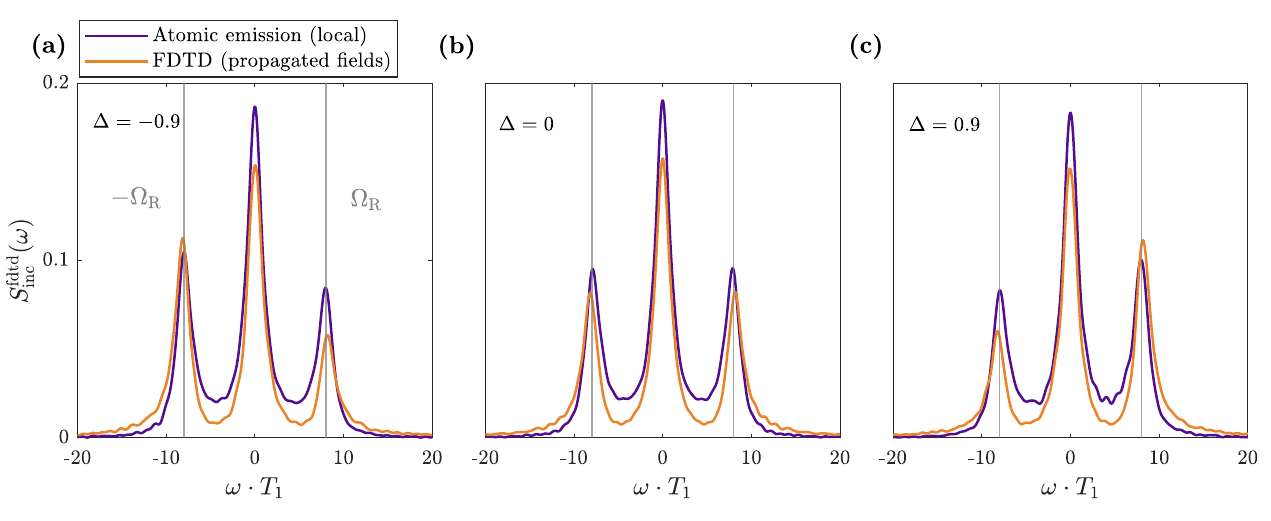}
    \caption{Comparison of the Mollow triplet under fixed Rabi driving, obtained from stochastic differential equations (at the atom, purple line) and from the propagated electric fields using the finite-difference time-domain (FDTD, orange line) method. The gray lines indicate the Rabi frequency $\Omega_{\mathrm{R}}$. Panels (a)–(c) show the Mollow triplet for slightly negative, zero, and positive detuning, respectively. Here, we measure time in units of the relaxation time $T_{1}$ and frequency in units of $1/T_{1}$ and set $T_{1}=1$.}
    \label{fig:FDTD_plot}
\end{figure*}
\subsection{Steady-state fluctuations and the Mollow spectrum}\label{sec:steady-state fluctuations}
In the driven dissipative regime, the steady state of the TLS is subject to intrinsic quantum fluctuations. These manifest in time-resolved correlation functions and ultimately shape the emission spectrum~\cite{meystre2007elements}. In particular, the incoherent spectrum encodes deviations from the mean field due to quantum noise and is defined as the steady-state correlation of the fluctuation operator $\delta\sigma_{\pm} = \sigma_{\pm} - \langle \sigma_{\pm} \rangle_{\mathrm{s.s.}}$~\cite{carmichael1993quantum}.
Using the trajectories of our stochastic Bloch vectors (Eq.~\eqref{eq:SDE_physics_style}), we construct the steady-state fluctuation correlation matrix by
\begin{equation}\label{eq:stochastic_steady_state_correlation}
\mathcal{C}_{ij}^{\mathrm{sto}}(\tau) = \overline{(  s_{1,i}(\tau) \,  s_{2,j}(0)  )}_{\mathrm{s.s.}} - \overline{(  s_{1,i}) }_{\mathrm{s.s.}}\, \overline{(s_{2,j} )}_{\mathrm{s.s.}}\,,
\end{equation}
where $i,j\in\{-,+,z\}$.
Taking the real part of the one-sided Fourier transform of $\mathcal{C}^{\mathrm{sto}}_{+-}(\tau)$, 
\begin{equation}\label{eq:mollow_spectrum_incoherent_first}
    S^{\mathrm{sto}}_{\mathrm{inc}}(\omega) \propto \Re \left\{ \int_{0}^{\infty} \mathrm{d}\tau \, \mathrm{e}^{-\mathrm{i}\omega\tau} \mathcal{C}^{\mathrm{sto}}_{+-}(\tau)\right\}\,,
\end{equation}
provides the incoherent spectrum, namely the Mollow triplet, a hallmark of resonance fluorescence. Since the spectrum is evaluated in the rotating frame of the laser frequency $ \omega_{\mathrm{L}} $, the variable $\omega $ represents the detuning from the drive. The coherent part is given by
\begin{equation}
    S^{\mathrm{sto}}_{\mathrm{coh}}(\omega) = \pi \,\overline{ ( s_{1,+}s_{2,-}  )}_{\mathrm{s.s.}} \delta(\omega) \,,
\end{equation}
such that the total emission spectrum reads as
\begin{equation}
     S^{\mathrm{sto}}(\omega) =  S^{\mathrm{sto}}_{\mathrm{inc}} + S^{\mathrm{sto}}_{\mathrm{coh}}\,.
\end{equation}
Figure~\ref{fig:theoretical_Mollow_spectra_colormap} shows the Mollow spectrum obtained from our stochastic approach, while systematically varying key parameters including the detuning (a) and driving strength (d). More precisely, Fig.~\ref{fig:theoretical_Mollow_spectra_colormap}(a) shows the total emission for a detuning ranging from \SIrange{-60}{60}{\micro\electronvolt}, corresponding to values up to twice the Rabi energy (\SI{30}{\micro\electronvolt}) and therefore entering the large detuning regime. The Mollow triplet remains clearly resolved across the entire range, with sideband positions shifting and intensities evolving with detuning. In (d), we are on resonance (zero detuning) and increase the Rabi energy, leading to a clear separation of the Mollow triplet sidebands, while the central peak remains the dominant spectral feature. The stochastic nature of the simulation approach is revealed by the slight noise in (a) and (d). This noise is directly related to the number of realizations of the stochastic Bloch vector, in the following denoted as \enquote{walkers}. 

\begin{figure}
    \centering    
    \scalebox{1.41}{
\begin{tikzpicture}[every node/.style={font=\normalsize}, 
  background rectangle/.style={fill=yellow!10, rounded corners=3pt}, 
  show background rectangle]

% Nodes
\node[draw=black, fill=white] (Liouvillian) at(0,0) {\large Liouvillian superoperator $\mathcal{L}$};
\node[draw=black, fill=white](Rho) at (-2.08,-1.25) {$\superket{\rho_{\mathrm{s.s.}}}$};
\node[draw=black, fill=white] (Cumulant) at (-1.18,-2.5) {$\bm{M}$};
\node[draw=black, fill=white] (Drift) at (0.2,-2.5) {\large $\bm{A}$};
\node[draw=black, fill=white] (Diffusion) at (2.4,-2.5) {\large $\bm{D}$};
\node[draw=black, fill=white] (Steadystate) at (0,-4) {\, \, \large Steady-state fluctuations $\mathcal{C}$\,\,};

% qrt
\draw[->, thick, draw=blue!40!red!95!black] (-1.52,-1.2) -- (-1.2,-1.2) -- (-1.2,-2.27);
\draw[->,thick, draw=blue!40!red!85!black] (-2.1,-1.54) -- (-2.1,-3.7);
\draw[->,thick,draw=blue!40!red!85!black] (-2.1,-0.3) -- (-2.1,-0.96);
\node[ draw=none ]  at (-1.7,-0.6) {$\mathrm{e}^{\mathcal{L}t}$};
\node[ draw=none ] at (-2.4,-2.5) {$\mathrm{qrt}$};

% grn
\draw[->,thick,draw=orange!70!black ] (0.2,-0.3) -- (0.2,-2.24);
\draw[->,thick,draw=orange!70!black] (0.2,-2.76) -- (0.2,-3.7);
\draw[->,thick,draw=orange!70!black] (-0.88,-2.65) -- (-0.09,-2.65);
\node[ draw=none ]  at (1.3,-1.3) {{$\bm{F} \bm{P}_{tl} \mathcal{L}^\dagger \bm{P}_{tl} \bm{F}^+$}};
\node[ draw=none ]  at (-0.1,-3.2) {$\mathrm{grn}$};

% sto
\draw[->,thick] (-0.88,-2.35) -- (-0.09,-2.35);
\draw[->,thick] (0.48,-2.49) -- (2.1,-2.49);
\draw[->,thick] (2.4,-2.76) -- (2.4,-3.7);
\node[ draw=none ]  at (2.1,-3.15) {$\mathrm{sto}$};

\end{tikzpicture} }

    \caption{Outline of the derivation of the steady-state fluctuations. Purple line: the steady-state solution in Liouville space is obtained via time propagation using $\mathrm{e}^{\mathcal{L}t}$. From this, the second-order cumulant matrix $\bm{M}$ is derived. Additionally, by applying the quantum regression theorem, we directly obtain the steady-state fluctuations $\mathcal{C}$. Orange line: By taking the traceless part of the adjoint Liouvillian $\mathcal{L}^{\dagger}$ in the computational basis, the drift matrix $\bm{A}$ is obtained. Then, applying the matrix exponential $\mathrm{e}^{\bm{A}t}$ to the second-order cumulant matrix $\bm{M}$ yields the steady-state fluctuations $\mathcal{C}$. Black line: The noise cross-covariance matrix $\bm{D}$ is obtained as a linear combination of $\bm{M}$ and $\bm{A}$. Factorizing $\bm{D}$ via singular value decomposition yields the noise matrices, which enable the construction of stochastic differential equations, leading to the steady-state fluctuations $\mathcal{C}$.}
    \label{fig:outline_steady_state_fluctuations}
\end{figure}
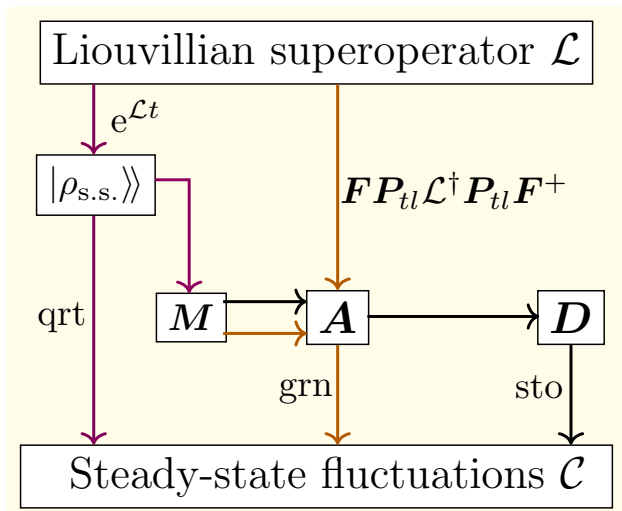

In order to quantify the influence of the number of walkers, we show in Fig.~\ref{fig:theoretical_Mollow_spectra_lineplot}(a)-(f) that the stochastic approach (sto) converges to the established solution. This is done by comparing the Mollow spectra, where the underlying steady-state fluctuations are either obtained by use of the quantum regression theorem (qrt)~\cite{lax1963formal} or the Green’s propagator method (grn) (Eq.~\eqref{eq:greens_propagator}), see Fig.~\ref{fig:outline_steady_state_fluctuations}. The steady-state fluctuations with the quantum regression theorem in Liouville space are given by~\cite{gyamfi2020fundamentals}
\begin{align}\label{eq:quantum_regression_theorem}
    \mathcal{C}^{\mathrm{qrt}}_{ij}(\tau) 
& = \superskp{\sigma_{i}(\tau)^{\dagger}}{\sigma_{j}(0) \rho_{\mathrm{s.s.}}}  -\superskp{\sigma_{i}^\dagger}{\rho_{\mathrm{s.s.}}}\superskp{\sigma_{j}^\dagger}{\rho_{\mathrm{s.s.}}} \nonumber \\ 
& =\sum^{4}_{k=1}\,\mathrm{e}^{\lambda_k \tau}
\, \superskp{ \sigma^{\dagger}_{i} }{r_k}
\, \superskp{ l_k }{ \sigma_{j} \rho_{\mathrm{s.s.}} } \nonumber \\ 
&- \superskp{\sigma_{i}^\dagger}{\rho_{\mathrm{s.s.}}}\superskp{\sigma_{j}^\dagger}{\rho_{\mathrm{s.s.}}} \,. 
\end{align}
For consistency, the steady-state fluctuations are also obtained directly from the drift matrix, the deterministic core of the SDE. We can project the adjoint Liouvillian $\mathcal{L^{\dagger}}$ onto the pseudospin basis to obtain a compact Bloch vector equation for the dynamics of single-time expectation values, see Methods. This yields a drift matrix $\bm{A}$ and an inhomogeneous term $\bm{b}$, allowing us to recast the quantum regression theorem into a matrix propagation form 
\begin{align}\label{eq:greens_propagator}
   \mathcal{C}^{\mathrm{grn}}_{ij}(\tau) &= \superskp{\sigma_{i}(\tau)^{\dagger}}{\sigma_{l}(0) \rho_{\mathrm{s.s.}}} -\superskp{\sigma_{i}^{\dagger}}{\rho_{\mathrm{s.s.}}} \superskp{\sigma_{l}^{\dagger}}{\rho_{\mathrm{s.s.}}} \nonumber \\
   & = \sum^{4}_{k=1}\, \bigl( \mathrm{e}^{\bm{A}\tau}\bigr)_{ik} M_{kj}\,.
\end{align}
In this representation, the time evolution of fluctuations becomes a simple matrix exponential acting on the cumulant matrix $\bm{M}$. Due to this structural similarity to linear response propagators, we refer to this as the Green’s (grn) propagator approach. The comparison of the detuned and power-dependent spectra in Fig.~\ref{fig:theoretical_Mollow_spectra_colormap} (a,d) $S^{\mathrm{sto}}(\omega)$, (b,e) $S^{\mathrm{grn}}(\omega)$, and (c,f) $S^{\mathrm{qrt}}(\omega)$, shows excellent agreement. Turning to Fig.~\ref{fig:theoretical_Mollow_spectra_lineplot}, the comparison between panels (a)–(c), which use a relatively low number of walkers ($N = \num{4e3}$), shows that the overall behavior of the Mollow spectrum is captured quite well. Panels (d)–(f) use a larger number of walkers ($N = \num[print-unity-mantissa=false]{1e5}$). This comparison demonstrates that the results obtained from our stochastic approach (sto) converge to the results from (grn) and (qrt).

We conclude this section by noting that in the stochastic approach, we do not explicitly fix an operator ordering during the derivation of the noise matrices $\bm{B}_{k}$ (Eq.~\eqref{eq:noise_matrices}) and the associated SDEs (Eq.~\eqref{eq:SDE_physics_style}). Instead, we construct the SDEs (Eq.~\eqref{eq:SDE_physics_style}) directly from the noise cross-covariance structure $\bm{D}$ (Eq.~\eqref{eq:noise_cross_covariance}), without committing to a particular phase-space representation. This is in contrast to traditional methods, where the choice of ordering (e.g., normal, antinormal, symmetric) guides the mapping from quantum operators to classical stochastic variables. Nonetheless, the solutions to the resulting SDEs must be interpreted with an ordering consistent with the physical observables of the problem to correctly reproduce quantum correlations.
\begin{figure*}
    \centering
    \hspace*{-1.6cm}
    \includegraphics[scale=0.94]{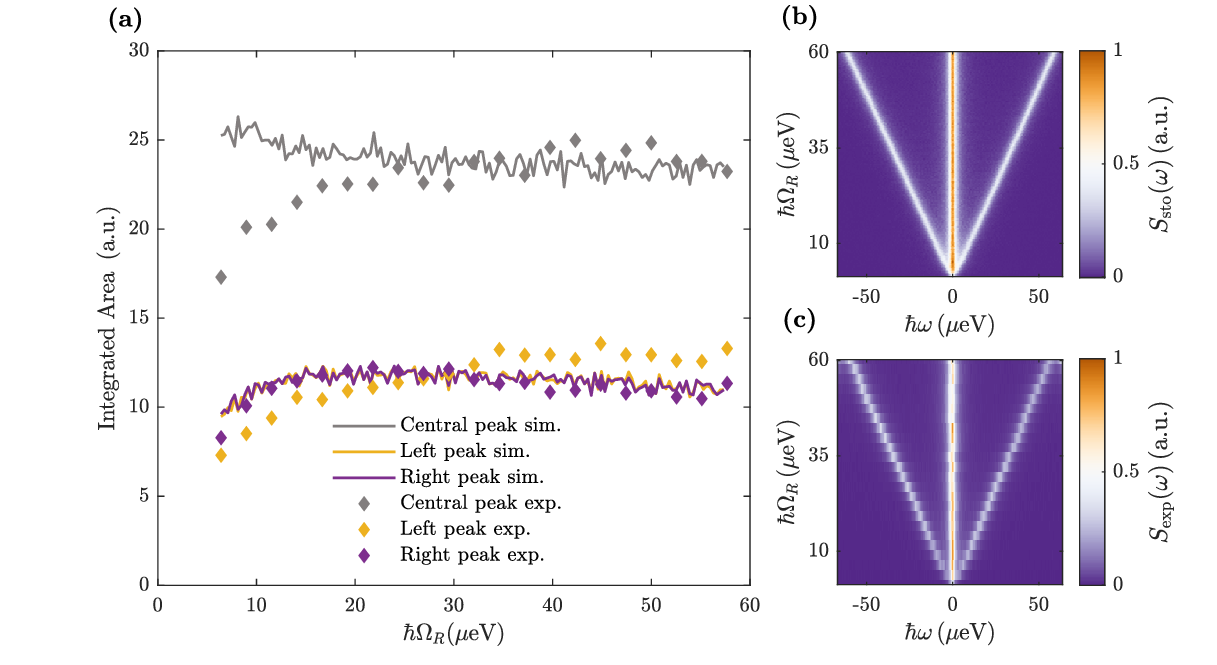}
    \caption{Power-dependent Mollow spectrum. Comparison of stochastic approach with experimental data obtained from InGaAs quantum dot. (a) Integrated peak areas of the Mollow spectrum, with solid lines corresponding to the stochastic approach and diamonds indicating the experimental results. (b) Power-dependent Mollow spectra (normalized) obtained by stochastic approach. (c) Experimental power-dependent Mollow spectra (normalized). Simulation parameters: Table~\ref{Simulation parameters}.}
    \label{fig:power_dependent_Mollow_spectrum}
\end{figure*}
\subsection{Stochastic Bloch vector coupled to a FDTD solver}\label{sec:stochastic_Bloch_vector_FDTD}

In the next step, we show that the trajectories of the stochastic Bloch vector can be coupled to a Maxwell solver. Here, we use the finite-difference time-domain (FDTD) method~\cite{yee1966numerical,taflove2005computational}, which allows us to directly propagate the electric and magnetic field in time. More precisely, we couple the stochastic Bloch vector components responsible for lowering and raising energy $s_{\mp}(t)$ to an electromagnetic source via% a spatial Gaussian
\begin{equation}\label{eq:source_current}
    \mathscr{J}^{\pm}_{\mathrm{source}}(x,t) 
    \;=\;  
    %\underbrace{
    \frac{1}{\sigma\sqrt{2\pi}}\,\mathrm{e}^{-\frac{(x-\mu)^{2}}{2\sigma^{2}}}%}_{\approx\delta(x)\,\text{for } \mu = 0,\, \sigma^{2}\ll1} 
    \;\mathrm{e}^{\pm \mathrm{i}\omega_\mathrm{c}t}\;s_{\mp}(t)\,,
\end{equation}
where $s_{-}(t) : =s_{1,-}(t)\,,\,\,s_{+}(t) : =s_{2,+}(t)$ and $\mathrm{e}^{\pm\mathrm{i}\omega_\mathrm{c}t}$ is the carrier-frequency phase factor. % We choose $\sigma^{2}\ll1$ such that the Gaussian effectively approximates $\delta(x)$ while remaining numerically stable.
For demonstration purposes, we use the source term (Eq.~\eqref{eq:source_current}) in a one‐dimensional FDTD solver with first‐order Mur boundary conditions. We propagate the electric and magnetic field by~\cite{inan2011numerical}
\begin{align}\label{eq:1D-FDTD_E-field}
 \left. \mathscr{E}_{y}^{(\pm)} \vphantom{\big|^{n+1}_{i}} \right|^{n+1}_{i}&=\left. \mathscr{E}_{y}^{(\pm)} \vphantom{\big|^{n_{i}}} \right|^{n}_{i}
 - \frac{\Delta t }{\varepsilon \Delta x} \biggl(
  \left. \mathscr{H}_{z}^{(\pm)} \vphantom{\big|^{n+\frac{1}{2}}_{i +\frac{1}{2}}} \right|^{n+\frac{1}{2}}_{i +\frac{1}{2}}
  -
  \left. \mathscr{H}_{z}^{(\pm)} \vphantom{\big|^{n+\frac{1}{2}}_{i -\frac{1}{2}}} \right|^{n+\frac{1}{2}}_{i -\frac{1}{2}}
 \biggr) \nonumber \\
 & - \frac{\Delta t}{\varepsilon}
 \left. \mathscr{J}^{(\pm)}_{\mathrm{source}} \vphantom{\big|^{n+\frac{1}{2}}_{i}} \right|^{n+\frac{1}{2}}_{i}\,,
\end{align}
\begin{equation}\label{eq:1D-FDTD_H-field}
\left. \mathscr{H}_{z}^{(\pm)} \vphantom{\big|^{n+\frac{1}{2}}_{i+\frac{1}{2}}} \right|^{n+\frac{1}{2}}_{i+\frac{1}{2}} 
= 
\left. \mathscr{H}_{z}^{(\pm)} \vphantom{\big|^{n-\frac{1}{2}}_{i+\frac{1}{2}}} \right|^{n-\frac{1}{2}}_{i+\frac{1}{2}} 
- 
\frac{\Delta t }{\Delta x} \biggl(
    \left. \mathscr{E}_{y}^{(\pm)} \vphantom{\big|^{n}_{i+1}} \right|^{n}_{i+1}
    -
    \left. \mathscr{E}_{y}^{(\pm)} \vphantom{\big|^{n}_{i}} \right|^{n}_{i}
\biggr) \,,
\end{equation}
where $\varepsilon$ is the dielectric function. %In summary, we note that the TLS radiates into the classical grid via $\mathscr{J}^{\pm}_{\mathrm{source}}(x,t)$.
This implementation enables the investigation of how vacuum‐induced phenomena and incoherent emission manifest in nanophotonic structures. We now reformulate the incoherent part of the optical spectrum in terms of the positive- and negative-frequency electric field components (denoted by superscripts $\pm$),
\begin{align}\label{eq:incoherent_Mollow_FDTD}
    S^{\mathrm{FDTD}}_{\mathrm{inc}} &\propto \mathrm{Re}\biggl\{ \int_{0}^{\infty} \mathrm{d}\tau\,\mathrm{e}^{-\mathrm{i}\omega\tau} \nonumber\\ &\times\Bigl( \overline{\mathscr{E}^{-}(\tau) \mathscr{E}^{+}(0)}_{\mathrm{s.s.}} 
    - \overline{ \mathscr{E}^{-}}_{\mathrm{s.s.}}\,  \overline{\mathscr{E}^{+}}_{\mathrm{s.s.}} \Bigr) \biggr\} \,,
\end{align}
where we average over different stochastic realizations of the electric field. In Fig.~\ref{fig:FDTD_plot}, we plot the results of Eq.~\eqref{eq:incoherent_Mollow_FDTD} together with those of Eq.~\eqref{eq:mollow_spectrum_incoherent_first} for a fixed Rabi driving and three different detuning scenarios. This provides a comparison between the Mollow triplet of the atomic emission (at the atom) and that obtained from the FDTD-propagated electric fields, both showing the expected spectral features.% Showing overall good agreement, the discrepancy in Fig.~\ref{fig:FDTD_plot} is most likely due to the replacement of the $\delta$-like quantum emitter by the Gaussian profile used in FDTD to comply with the spatial grid. We conclude that this does not affect the qualitative agreement between the two approaches. 
\begin{figure*}
    \centering
    \hspace*{-1.6cm}
    \includegraphics[scale=0.94]{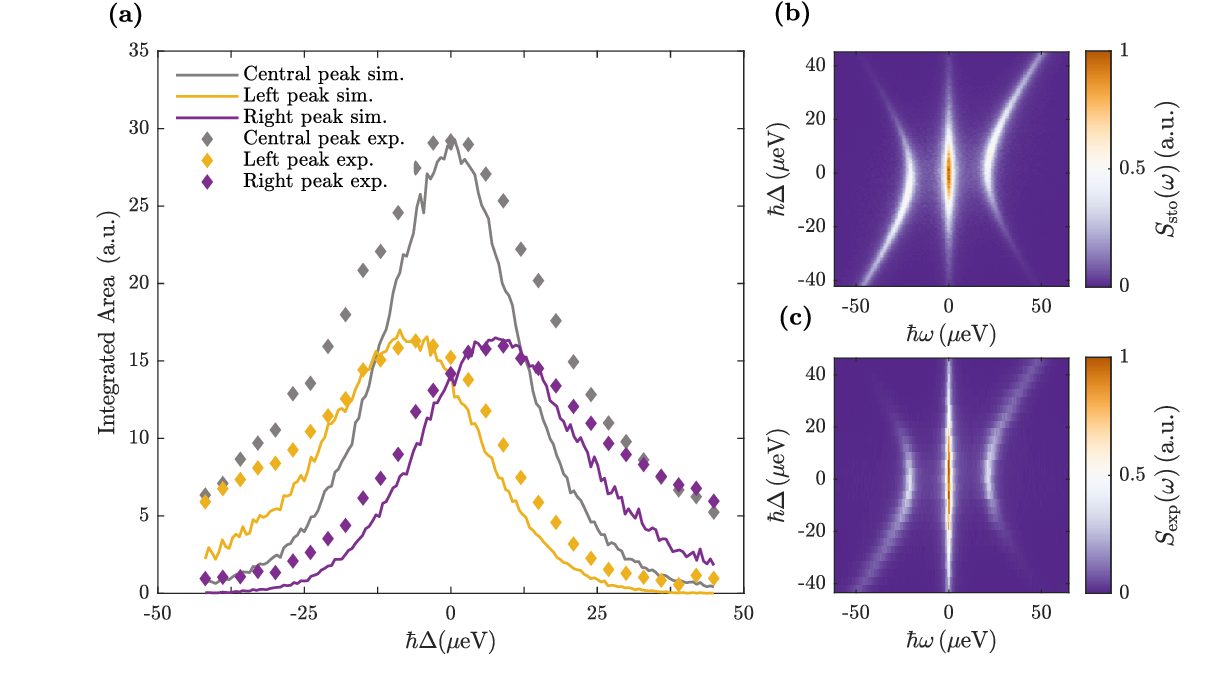}
    \caption{Detuning-dependent Mollow spectrum. Comparison of stochastic approach with experimental data obtained from InGaAs quantum dot. (a) Integrated peak areas of the Mollow spectrum, with solid lines corresponding to the stochastic approach and diamonds indicating the experimental results. (b) Detuning-dependent Mollow spectra (normalized) obtained by stochastic approach. (c) Experimental detuning-dependent Mollow spectra (normalized). Simulation parameters: Table~\ref{Simulation parameters}.}
    \label{fig:detuning_dependent_Mollow_spectrum}
\end{figure*}
\subsection{Emission spectrum of an InGaAs quantum dot}\label{sec:comparison_InGaAs_quantum_dot}
For model validation, the simulation results are compared to experimental measurements on an InGaAs quantum dot, where the negatively charged exciton serves as a TLS. For the measurement, the sample is cooled to \SI{4.2}{\kelvin} and integrated into a resonance fluorescence setup to separate the laser signal from the QD emission by cross-polarized filtering. The emission is recorded with a high-resolution scanning Fabry-Perot interferometer. Using our stochastic approach, we successfully reproduce the power-dependent emission spectrum of a monochromatic resonantly driven QD for moderate Rabi energies. The model also captures the behavior of the QD emission under varying laser detuning.

We first investigate how the emission spectrum evolves with increasing laser power. A continuous wave laser tuned in resonance with the QD transition at the energy $E=\SI{1362.0315}{\milli\electronvolt}$ is used, and the excitation power is increased from \SI{1}{\nano\watt} to \SI{2.21}{\micro\watt} in quadratically increasing steps. In Fig.~\ref{fig:power_dependent_Mollow_spectrum}(c), we plot the experimental data against the Rabi energy $\hbar\Omega_{\mathrm{R}}=\sqrt{\hbar\,\eta_{\mathrm{R}}P_{\mathrm{exc}}}$ and the relative photon energy $\hbar \omega$, where $\eta_{\mathrm{R}}$ is a dimensionless scaling factor. For comparison, we plot our stochastic approach in Fig.~\ref{fig:power_dependent_Mollow_spectrum}(b), whereby the stochasticity is reflected in the slight noise in the data. To obtain a more quantitative comparison, we fit the spectra for each $\hbar\Omega_{\mathrm{R}}$ with Lorentzian line shapes and integrate the area under each peak (see Fig.~\ref{fig:power_dependent_Mollow_spectrum}(a)). The onset of Mollow triplet formation is observed for a Rabi energy $\hbar\Omega_{\mathrm{R}}>\SI{6.4}{\micro\electronvolt}$. For small Rabi energies, the peak areas extracted from experiment and simulation differ slightly, which we mainly attribute to difficulties in disentangling the contributions of the overlapping peaks. 
For characterizing the off-resonant behavior of the QD, we fix the driving strength at $\hbar\Omega_{\mathrm{R}} = \SI{21}{\micro\electronvolt}$, and detune the laser in
equidistant steps from \SIrange{-42}{45}{\micro\electronvolt} with respect to the QD $X^{-}$ resonance. In Fig.~\ref{fig:detuning_dependent_Mollow_spectrum}(b),(c), we show the comparison of the experimental data and our stochastic approach. For each $\hbar\Delta$, we characterize the resulting spectrum by integrating the area under the fitted peaks, see Fig.~\ref{fig:detuning_dependent_Mollow_spectrum}(a). The different peak widths between experiment and simulation might be due to additional dephasing channels in the experiment or some energy jitter during measurement acquisition.

%%%%%%%%%%%%%%%%%%%%%%%%%%%%%%%%%%%%%%%%%%%%%%%%%%%%%%%%%%%%%%%%%%%%%%%%%%%%%%%%%%%%%%%%%%%%%%%%%%%%%%%%%%%%%%%%%%%%%%%%%%%%%%%%%%%%DISCUSSION/OUTLOOK%%%%%%%%%%%%%%%%%%%%%%%%%%%%%%%%%%%%%%%%%%%%%%%%%%%%%%%%%%%%%%%%%%%%%%%%%%%%%%%%%%%%%%%%%%%%%%%%%%%%%%%%%%%%%%%%%%%%%%%%%%%%%%%%%%%%%%%%%%

\section{Discussion\label{sec:discussion}}

We demonstrate that allowing a non-symmetric diffusion matrix in the derivation of SDEs    fills a critical gap in the modeling of non-trivial noise processes. As a quantum optical example showcasing the benefits of our approach, we accurately reproduce the power- and detuning-dependent emission spectrum of an InGaAs QD. Here, the resulting smooth transition from quantum to classical regimes guarantees compatibility with classical electromagnetic solvers.

The core of our formulation is an SVD of the non-symmetric diffusion matrix, which yields a doubled system of SDEs. The controlled distribution of noise contributions across these SDEs facilitates the encoding of correlated fluctuations that symmetric-diffusion models cannot represent. Conceptually, the non-symmetric diffusion matrix captures the consequences of operator non-commutativity in quantum mechanics. For the case of a TLS, which in our work is realized by an InGaAs QD, the system of SDEs may be viewed as a stochastic Bloch-vector description whose solutions can be injected into electromagnetic solvers via a source-current coupling. In doing so, it reproduces experimentally relevant spectral features, while retaining a possible path to device scale modeling through the EM-coupling.

This is of particular interest because many Maxwell-Bloch solvers either neglect or approximate vacuum-induced fluctuations and therefore struggle to reproduce the incoherent part of the spectrum of a quantum emitter~\cite{allen1987optical,jirauschek2019optoelectronic,popp2024modeling,pusch2012coherent,andreasen2010numerical,slavcheva2004fdtd}. We are now able to naturally supply FDTD solvers with the missing quantum noise, which enables quantitative predictions of emission spectra in linear photonic environments without back action. While we here focus on the steady-state dynamics, the stochastic framework itself is not limited to steady-state conditions. Although our interface with FDTD does not yield a fully self-consistent simulation that includes the back-action on the quantum emitter, it can nevertheless model linear photonic environments that act as passive filters, such as optical cavities that transmit certain wavelengths and waveguides designed to pass specific modes.

Compared to prior stochastic approaches in quantum optics, most of which target ensembles of TLSs~\cite{carmichael2008statistical,drummond1991quantum}, our approach addresses a single TLS without any approximations. Moreover, we anticipate no major obstacles standing in the way of an extension to multi-level systems using generalized Bloch vectors~\cite{hioe1981n-level,slavcheva2002coupled,jirauschek2019optoelectronic}. 
   
Our derivation is currently established for two-time correlations. Many key observables in quantum optics fall into this category, and even second-order coherence can be handled when the required averages are recast in terms of two-time correlations~\cite{meystre2007elements}. Still, the extension to multi-time correlations and non-Gaussian noise processes is an attractive goal, since it enables the modeling of single photon emission~\cite{jezek2011experimental,fiurasek2021quantum}. Finally, despite the fact that our approach removes operator ordering on quantum side, it relocates the choice to the classical post-processing stage. One must adopt a consistent ordering when mapping stochastic variables to observables. Notably, recent research has focused on filtering out the coherent and incoherent components of emission spectra, a task that our framework could support~\cite{kim2024unlocking, lopez2018joint, hanschke2020origin}. 

In conclusion, our approach offers a practical route to quantitatively link quantum noise and classical fields, while also providing a general methodology for tailoring correlated noise in stochastic models beyond quantum optics.

%%%%%%%%%%%%%%%%%%%%%%%%%%%%%%%%%%%%%%%%%%%%%%%%%%%%%%%%%%%%%%%%%%%%%%%%%%%%%%%%%%%%%%%%%%%%%%%%%%%%%%%%%%%%%%%%%%%%%%%%%%%%%%%%%%%%%%%%%%%%%%%%%APPENDIX%%%%%%%%%%%%%%%%%%%%%%%%%%%%%%%%%%%%%%%%%%%%%%%%%%%%%%%%%%%%%%%%%%%%%%%%%%%%%%%%%%%%%%%%%%%%%%%%%%%%%%%%%%%%%%%%%%%%%%%%%%%%%%%%%%%%%%%%%%%%%%%%%%%%%%

\begin{center}
    \rule{0.8\linewidth}{0.6pt}
\end{center}

\section{Methods}\label{sec:methods}
We adopt the following notation: $(\cdot)^{\mathrm{T}}$ denotes the transpose, $(\cdot)^{*}$ denotes the complex conjugate, $(\cdot)^{\dagger}=((\cdot)^{*})^{\mathrm{T}}$ denotes the Hermitian adjoint, $\langle \cdot \rangle$ denotes the expectation value as used in quantum theory and $\overline{(\cdot)}$ denotes the sample mean. 

\subsection{Pseudospin operators}\label{sec:pseudospin_operators_methods}

The following convention is used for the energy eigenstates of the TLS (ground state $\ket{g}$ and excited state $\ket{e}$):
\begin{equation}
\ket{g} = \begin{pmatrix}
        0 \\
        1 \\
    \end{pmatrix} \,, \quad
\ket{e} = \begin{pmatrix}
        1 \\
        0 \\
    \end{pmatrix} .
\end{equation}
We write the pseudospin operators as outer products regarding our chosen basis $\sigma_{ij}=\ketbra{i}{j}$ with $i,j\in\{e,g\}$, which are elements of the space $\mathcal{B(\mathbb{C}}^{2})$ of bounded linear operators on the Hilbert space $\mathbb{C}^{2}$. However, we index the pseudospin operators in a way that reflects their physical action, using $-, +, z$ instead of $e, g$.

\begin{table}[h!]
\centering
\label{tab:pseudospin_matrices_methods}
\caption{Matrix representation of pseudospin operators.}
\begin{tabularx}{\columnwidth}{|c|X|}
\hline
\textbf{Operator} & \textbf{Matrix representation} \\
\hline
Lowering $\sigma_{-}$ & \small$\displaystyle\sigma_{-}:=\ketbra{g}{e} = \begin{pmatrix}
        0 & 0 \\
        1 & 0 \\
    \end{pmatrix} $  \\

Raising $ \sigma_{+}$ & \small$\displaystyle\sigma_{+}:= \ketbra{e}{g} = \begin{pmatrix}
        0 & 1 \\
        0 & 0 \\
    \end{pmatrix} $  \\

Population difference $\sigma_{z}$ & \small$\displaystyle\sigma_{z}:=\ketbra{e}{e}-\ketbra{g}{g} = \begin{pmatrix}
        1 & 0 \\
        0 & -1 \\
    \end{pmatrix} $ \\

\hline
\end{tabularx}
\end{table}
These operators satisfy the commutation relations:
\begin{equation*}
    [\sigma_{z},\sigma_{\pm}] = \pm 2\sigma_{\pm}\,,\,\, [\sigma_{+},\sigma_{-}] = \sigma_{z}\, .
\end{equation*}

\subsection{Liouville space}\label{sec:Liouville_space_methods}

The formulation in Liouville space resembles Dirac’s bracket notation, and is primarily chosen for its readability and structural clarity. We work with the Hilbert space $\mathcal{H} = \mathbb{C}^2$ for a TLS, on which operators act as $2 \times 2$ complex matrices. The Liouville space $\mathbb{L}_{2}$ is defined as the vector space isomorphic to $\mathbb{C}^4$, obtained by vectorizing operators on $\mathcal{H}$. We adopt the following convention for vectorization of operators $\ketbra{a}{b} \in \mathcal{B}(\mathbb{C}^2)$ to Liouville space $\mathbb{L}_{2}$
\begin{equation}\label{eq:vectorization_methods}
   \mathrm{vec}\bigl( \ketbra{a}{b} \bigr) = \ket{a} \otimes \ket{b}^* = \superket{a,b}\,,
\end{equation}
where $\otimes$ denotes the Kronecker product, and $\superket{\cdot}$ indicates a superket, i.e., a vectorized operator in Liouville space. The corresponding superbra is defined as
\begin{equation}\label{eq:suberbra_methods}
   \superbra{a,b} := \bra{a} \otimes \bra{b}^*\,,
\end{equation}
so that the Hilbert-Schmidt inner product between operators $A$ and $B$ satisfies~\cite{gyamfi2020fundamentals}
\begin{equation}\label{eq:Hilbert_Schmidt_methods}
   \superskp{A}{B} := \mathrm{Tr}(A^\dagger B) = \mathrm{vec}(A)^\dagger \mathrm{vec}(B)\,.
\end{equation}
Under this convention, pseudospin operators, originally $2 \times 2$ matrices on the Hilbert space $\mathbb{C}^2$, are mapped to $4 \times 1$ vectors in Liouville space. Their vectorized forms are shown in Table~\ref{tab:superkets_pseudospin_methods}.
\begin{table}[h!]
\centering
\caption{Vectorization of pseudospin operators in Liouville space.}
\label{tab:superkets_pseudospin_methods}

\begin{tabularx}{\columnwidth}{|c|X|}
\hline
\textbf{Operator} & \textbf{Superket} \\
\hline
Lowering  $\sigma_{-}$ & \scriptsize$\displaystyle \superket{\sigma_-} = \mathrm{vec}(\ketbra{g}{e}) = \begin{pmatrix}0\\0\\1\\0\end{pmatrix}$ \\
Raising $ \sigma_{+}$ & \scriptsize$\displaystyle \superket{\sigma_+} = \mathrm{vec}(\ketbra{e}{g}) = \begin{pmatrix}0\\1\\0\\0\end{pmatrix}$ \\
Population difference $\sigma_{z}$ & \scriptsize$\displaystyle \superket{\sigma_z} = \mathrm{vec}(\ketbra{e}{e}-\ketbra{g}{g}) = \begin{pmatrix}1\\0\\0\\-1\end{pmatrix}$ \\
Ground state $\sigma_{\mathrm{gg}}$ & \scriptsize$\displaystyle \superket{\sigma_{\mathrm{gg}}} = \mathrm{vec}(\ketbra{g}{g}) = \begin{pmatrix}0\\0\\0\\1\end{pmatrix}$ \\
\hline
\end{tabularx}
\end{table}
The vectorization map $\mathrm{vec}: \mathcal{B}(\mathbb{C}^2) \to \mathbb{L}_{2}$ is linear and preserves the Hilbert-Schmidt inner product, making it a useful tool to translate operator equations into vector equations. Additionally, we define superoperators as elements of the space $\mathcal{B}(\mathbb{L}_{2})$, the set of bounded linear operators acting on the Liouville space $\mathbb{L}_{2}$. These superoperators are represented by $4 \times 4$ complex matrices. For example, the Liouvillian superoperator $\mathcal{L} \in \mathbb{C}^{4 \times 4}$ governs the dynamics of density operators in Liouville space. The super(anti)commutator maps $ \llbracket\cdot,\cdot \rrbracket: \mathcal{B}(\mathbb{C}^2)\times\mathcal{B}(\mathbb{C}^2) \to \mathcal{B}(\mathbb{L}_{2}) $, so for $X,\, Y \in  \mathcal{B}(\mathbb{C}^2)$, the supercommutator is defined as~\cite{graham2018kronecker}
\begin{equation}\label{eq:supercommutator_methods}
    \llbracket X ,Y \rrbracket = X \otimes Y^{\mathrm{T}} - Y \otimes X^{\mathrm{T}}\,,
\end{equation}
and the superanticommutator is given by
\begin{equation}\label{eq:superanticommutator_methods}
    \llbracket X ,Y \rrbracket_{+} = X \otimes Y^{\mathrm{T}} + Y \otimes X^{\mathrm{T}} \,.
\end{equation}
To obtain expectation values and second-order moments for the vectorized pseudospin operators, we can adopt the formalism outlined above, yielding
\begin{align}\label{eq:expectation_value_superskp_methods}
     \langle \sigma_{i}\rangle_{\mathrm{s.s.}} &= \mathrm{Tr}( \sigma_{i}  \rho_{\mathrm{s.s.}}  ) \nonumber\\
     &= \mathrm{Tr}\bigl( (\sigma_{i}^{\dagger})^{\dagger}  \rho_{\mathrm{s.s.}}  \bigr)=   \superskp{\sigma_{i}^{\dagger}}{\rho_{\mathrm{s.s.}}}
\end{align}
and
\begin{align}\label{eq:second_moments_superskp_methods}
    \langle \sigma_{i}\sigma_{j}\rangle_{\mathrm{s.s.}} &= \mathrm{Tr}( \sigma_{i}\sigma_{j}\rho_{\mathrm{s.s.}} ) \nonumber \\ 
    &= \mathrm{Tr}\bigl( \bigl((\sigma_{i}\sigma_{j})^{\dagger}\bigr)^{\dagger} \rho_{\mathrm{s.s.}} \bigr)=\superskp{(\sigma_{i}\sigma_{j})^{\dagger}}{\rho_{\mathrm{s.s.}}} \,.
\end{align}

\subsection{Second-order cumulant matrix}\label{sec:second_order_cumulant_methods}

The Lindblad master equation in its typical form reads~\cite{lindblad1976generators,breuer2002theory} 
\begin{equation}\label{eq:LME_rho_methods}
    \partial_{t}\rho = -\mathrm{i} \hbar^{-1} [ H,\rho ] + \sum_{k} \Bigl( L_{k} \rho L_{k}^{\dagger} - \frac{1}{2}\{L_{k}^{\dagger} L_{k} \rho\} \Bigr )\, ,
\end{equation}
where $L_{k}$ are the Lindblad jump operators.
Here, we apply the Lindblad master equation approach in a two-dimensional Liouville space $\mathbb{L}_{2}$. The Liouvillian superoperator $\mathcal{L}$ for a TLS in the Schrödinger picture can be expressed through
\begin{equation}\label{eq:Liouvillian_methods}
    \mathcal{L} = -\mathrm{i}\hbar^{-1} \llbracket H, \mathds{1}_{2} \rrbracket + \sum_{k=1}^{2} \Bigl( L_{k} \otimes L_{k}^{*} - \frac{1}{2}\llbracket  L_{k}^{\dagger}L_{k}, \mathds{1}_{2}\rrbracket_{+} \Bigr) \,,
\end{equation}
where $\mathds{1}_{2}$ is the two-dimensional identity operator. In our setting, the Lindblad jump operators are defined as $L_{1} = \sqrt{\Gamma_1} \sigma_{-}$ and $L_{2} = \sqrt{\Gamma_2}{(\mathds{1}_{2}+\sigma_{z})}/{2}$, where $\Gamma_1 = 1/T_1$ and $\Gamma_2 = 2/T_2$. Here, $T_1$ denotes the relaxation time, while $T_2$ is the phenomenological dephasing time that accounts for processes such as phonon-induced dephasing~\cite{hohenester2020nano}. Plugging Eq.~\eqref{eq:Hamiltonian} and the Lindblad jump operators in Eq.~\eqref{eq:Liouvillian_methods} yields
\begingroup
\footnotesize
\begin{equation}\label{eq:eplicit_Liouville_superoperator_methods}
\mathcal{L} =
\begin{pmatrix} 
        -\Gamma_{1}  &  \frac{-\mathrm{i}\Omega_{\mathrm{R}}}{2} & \frac{\mathrm{i}\Omega_{\mathrm{R}}}{2} & 0 \\ 
       \frac{-\mathrm{i}\Omega_{\mathrm{\mathrm{R}}}}{2} & -\mathrm{i}\Delta-\frac{1}{2}\bigl(\Gamma_{1} + \Gamma_{2} \bigr) & 0 & \frac{\mathrm{i}\Omega_{\mathrm{\mathrm{R}}}}{2} \\
       \frac{\mathrm{i}\Omega_{\mathrm{\mathrm{R}}}}{2} & 0 & \mathrm{i}\Delta-\frac{1}{2}\bigl(\Gamma_{1} + \Gamma_{2} \bigr)  & \frac{-\mathrm{i}\Omega_{\mathrm{\mathrm{R}}}}{2} \\
        \Gamma_{1} & \frac{\mathrm{i}\Omega_{\mathrm{R}}}{2} & \frac{-\mathrm{i}\Omega_{\mathrm{R}}}{2} & 0
    \end{pmatrix} \,.
\end{equation}
\endgroup
The time‐independent Liouvillian superoperator (Eq.~\eqref{eq:Liouvillian_methods}) is known to be the generator of a quantum Markovian evolution, i.e., it generates a completely positive, trace‐preserving semigroup $ \{\mathrm{e}^{t\mathcal{L}}|{t\ge0}\} $ under the Born-Markov and secular approximations~\cite{rivas2012open,alicki2007quantum}.
The dynamics for the vectorized density matrix in Liouville space is then governed by
\begin{equation}\label{eq:time-evolution_methods}
    \superket{\rho(t)} = \mathrm{e}^{\mathcal{L}t} \superket{\rho(0)} \,.
\end{equation}
To compute this explicitly, we diagonalize $\mathcal{L}$ and expand the solution in terms of its biorthonormal eigenbasis $\{ \superket{r_k}, \superbra{l_k} \}$, yielding
\begin{equation}\label{eq:time-evolution_explicit_methods}
    \superket{\rho(t)} = \sum_{k=1}^{4} \mathrm{e}^{\lambda_k t} \superket{r_k}\superskp{l_k}{\rho(0)}\,,
\end{equation}
where $\lambda_k$ are the complex eigenvalues of $\mathcal{L}$, with corresponding right eigenvectors \(\superket{r_k}\) and left eigenvectors \(\superbra{l_k}\). We obtain the steady-state solutions in the long time limit $\superket{\rho_{\mathrm{s.s.}}}:= \lim_{t\to \infty} \superket{\rho(t)}$ via
\begin{equation}\label{eq:rho_steady_methods}
    \superket{\rho_{\mathrm{s.s.}}} = \lim_{t \to \infty} \sum_{k=1}^{4} \mathrm{e}^{\lambda_{k}t} \superket{r_{k}}\superskp{l_{k} }{\sigma_{\mathrm{gg}}}\,,
\end{equation}
where the ground-state supervector $\superket{\sigma_{\mathrm{gg}}}$, see Table~\ref{tab:superkets_pseudospin_methods}, is used as an initial state. We define the steady-state second-order cumulant matrix for the pseudospin operators as
\begin{equation}\label{es:second_order_cummulant_methods}
    \bm{M} = (M_{ij})_{i,j\in\{-,+,z\}} \in \mathbb{C}^{3\times3}\,,
\end{equation}
with entries 
\begin{equation} \label{eq:second_order_cumulant_entries_methods}
    M_{ij} = \superskp{(\sigma_{i}\sigma_{j})^{\dagger}}{\rho_{\mathrm{s.s.}}} - \superskp{\sigma_{i}^{\dagger}}{\rho_{\mathrm{s.s.}}} \superskp{\sigma_{j}^{\dagger}}{\rho_{\mathrm{s.s.}}} \,.
\end{equation}
In the Heisenberg picture, operators evolve under the adjoint Liouvillian as $\sigma_i(\tau) = \mathrm{e}^{\mathcal L^\dagger \tau}\sigma_{i}$. Using this, the quantum regression theorem in Liouville space can therefore be written as
\begin{align}\label{eq:qrt_notation_methods}
    \langle\sigma_{i}(\tau)\sigma_{j}(0)\rangle_{\mathrm{s.s.}} 
    & = \mathrm{Tr}( \sigma_{i}(\tau)\sigma_{j}(0)\rho_{\mathrm{s.s.}}) \nonumber\\
    & =\superskp{\sigma_{i}(\tau)^{\dagger}}{\sigma_{j}(0) \rho_{\mathrm{s.s.}}} \nonumber \\
    & = \superbra{\sigma_{i}^{\dagger}}\mathrm{e}^{\mathcal{L}\tau}\superket{\sigma_{j}(0) \rho_{\mathrm{s.s.}}} \nonumber \\
    & = \sum_{k=1}^{4} \mathrm{e}^{\lambda_{k}\tau}\superskp{\sigma_{i}^{\dagger}}{r_{k}} \superskp{l_{k}}{\sigma_{j}(0) \rho_{\mathrm{s.s.}}} \,.
\end{align}

\subsection{Projection of \texorpdfstring{$\mathcal{L}^{\dagger}$}{the adjoint Liouvillian}}\label{sec:projection adjoint Louillian methods}
We can project the adjoint Liouvillian $\mathcal{L}^{\dagger}$ by suitable matrix transformations into the computational basis, which yields the drift matrix $\bm{A}$ and inhomogeneity term $\bm{b}$. The drift matrix $\bm{A}$ is known to be the deterministic time evolution matrix for Heisenberg-Langevin equations~\cite{gardiner2004quantum}. For a TLS, it is given by the traceless part of $\mathcal{L}^\dagger$ in the computational basis. We define
\begin{equation}
    \bm{P}_{\mathrm{tl}} = \mathds{1}_4 - \frac{1}{2} \superket{\mathds{1}_{2}}\!\superbra{\mathds{1}_2}
\end{equation}
as the projection onto the traceless subspace. Furthermore, we define the matrix
\begin{equation}
    \bm{F} = \ket{0}\!\!\superbra{\sigma_-} + \ket{1}\!\!\superbra{\sigma_+} + \ket{2}\!\!\superbra{\sigma_z}
\end{equation}
that maps the basis of the traceless subspace $\{\superket{\sigma_-}, \superket{\sigma_+}, \superket{\sigma_z}\}$ into the three-dimensional computational basis $\{|0\rangle = (1,0,0)^{\mathrm{T}}, |1\rangle = (0,1,0)^{\mathrm{T}}, |2\rangle = (0,0,1)^{\mathrm{T}}\}$. Thus, $\bm{F}$ is a $3\times 4$ rectangular matrix.
Denoting the Moore-Penrose inverse \cite{penrose1955generalized} of $\bm{F}$ by $\bm{F}^+$, we have
\begin{equation}\label{eq:projection_adjoint_Liouvillian_methods}
    \bm{A} = \bm{F}\bm{P}_{\mathrm{tl}}\mathcal{L}^\dagger \bm{P}_{\mathrm{tl}}\bm{F}^+ \,,
\end{equation}
which yields
\begin{equation}\label{eq:explicit_drift_methods}
\bm{A} =
\begin{pmatrix}
   -\mathrm{i}\Delta-\frac{\Gamma_{1} + \Gamma_{2}}{2}  & 0 &  -\frac{\mathrm{i}\Omega_{\mathrm{R}}}{2}\\
    0 & \mathrm{i}\Delta - \frac{\Gamma_{1} + \Gamma_{2}}{2}  & \frac{\mathrm{i}\Omega_{\mathrm{R}}}{2} \\
   -\mathrm{i}\Omega_{\mathrm{R}} &  \mathrm{i}\Omega_{\mathrm{R}} & -\Gamma_{1}  
\end{pmatrix}\,.
\end{equation}
The inhomogeneity $\bm{b}$ is obtained by
\begin{equation}\label{eq:projection_inhomogeneity}
\bm{b} = \superbra{\mathds{1}_2}\bm{P}_\mathrm{t} \mathcal{L}^{\dagger}\bm{P}_{\mathrm{tl}}\bm{F}^{+}\,,
\end{equation}
where $\bm{P}_{\mathrm{t}}=\mathds{1}_4 - \bm{P}_{\mathrm{tl}}$ projects on the trace subspace, yielding
\begin{equation}\label{eq:inhomogeneity }
\bm{b} =
\begin{pmatrix} 
   0\\
    0  \\
   -\frac{1}{T_{1}}  
\end{pmatrix}\,.
\end{equation}
With $\bm{A}$ and $\bm{b}$ at hand, one can construct a Bloch vector equation for the expectation values of the pseudospin operators in the Heisenberg picture,
\begin{equation}\label{eq:bloch_vector_methods}
    \partial_{t}\bm{u} = \bm{A}\bm{u} + \bm{b}\,,
\end{equation}
where $\bm{u}= (\langle \sigma_{-}(t)\rangle, \langle \sigma_{+}(t)\rangle, \langle \sigma_{z}(t)\rangle )^{\mathrm{T}}$.

\noindent We can also apply the quantum regression theorem using the drift matrix $\bm{A}$. The entries of the steady-state fluctuations are given by
\begin{align}\label{eq:greens_propagator_methods}
   \mathcal{C}^{\mathrm{grn}}_{ij}(\tau) & = \sum^{4}_{k=1}\, \bigl( \mathrm{e}^{\bm{A}\tau}\bigr)_{ik} M_{kj}\,,
\end{align}
where the Green's propagator method is primarily used for benchmarking our stochastic approach.

\subsection{Stochastic Approach}\label{sec:stochastic_approach_methods}
We follow the standard theory of linear SDEs, matrix analysis, and Gaussian processes~\cite{horn2012matrix,pavliotis2014stochastic,vankampen2007stochastic,kloeden1992numerical}. The main modification is a non-symmetric diffusion matrix. The second-order cumulants do not depend on any particular operator ordering. The non-commutativity of the operators involved is responsible for the non-symmetric matrix $\bm{D}$. The equation of motion for the covariance matrix $\bm{C}$ is given by~\cite{carmichael1999statistical}
\begin{equation} \label{eq:EOM_covariance_methods}
    \partial_{t}\bm{C} = \bm{A}\bm{C} + \bm{C}  \bm{A}^{\mathrm{T}} +\bm{D}\,,
\end{equation}
where $\bm{A}$ and $\bm{D}$ are the drift and diffusion matrix, respectively. In the steady state, this gives the (time-continuous) Lyapunov equation
\begin{equation}\label{eq:algebraic_lyapunov}
- \bm{D} = \bm{A} \bm{C} + \bm{C}  \bm{A}^{\mathrm{T}} \,.
\end{equation}
If $\bm{D}\succeq 0$ and $\bm{A}$ is Hurwitz (i.e., all eigenvalues of $\bm{A}$ have strictly negative real parts), the unique steady-state solution $\bm{C}$ is positive semidefinite, ensuring that all variances (diagonal elements) are non-negative~\cite{pavliotis2014stochastic}.
Rather than solving Eq.~\eqref{eq:algebraic_lyapunov} for the covariance $\bm{C}$, we express the diffusion matrix $\bm{D}$ as a linear combination of the drift matrix $\bm{A}$ and the second-order cumulant $\bm{M}$ 
\begin{equation}
    - \bm{D} = \bm{A}  \bm{M} +  \bm{M}  \bm{A}^{\mathrm{T}} \,.
\end{equation}
This yields a complex, non-symmetric matrix $\bm{D}$, which no longer has the standard interpretation of a diffusion matrix in SDE theory. Rather, $\bm{D}$ is interpreted as the noise cross-covariance matrix. To factorize $\bm{D}$, we use the SVD Eq.~\eqref{eq:SVD}, and construct the corresponding complex noise matrices according to Eq.~\eqref{eq:noise_matrices}.
The noise matrices $\bm{B}_{1}$ and $\bm{B}_{2}$, together with the drift matrix $\bm{A}$, define two linear SDEs with additive noise in Itô form
\begin{equation}\label{eq:SDEs_methods}
\begin{cases}
\mathrm{d}\bm{X}_{k,t} = \left(\bm{A}\bm{X}_{k,t} + \bm{b}\right)\mathrm{d}t + \bm{B}_{k} \, \mathrm{d}\bm{W}(t), \\[6pt]
\bm{X}_{k,0} = \bm{x}_{k,0},
\end{cases}
\quad k \in \{1,2\}\,,
\end{equation}
where $\bm{X}_{k,t}$ is a stochastic process representing the system state, $\bm{b}$ is a constant inhomogeneous term, $\bm{W}(t)$ is a vector of independent real-valued Wiener processes, and $\bm{x}_{k,0}$ denotes the (possibly random) initial condition~\cite{pavliotis2014stochastic}.

Since $T_1,\,T_2,\,\Omega_{\mathrm{R}}>0$, the drift matrix $\bm{A}$ is invertible and Hurwitz, which ensures that the dynamics relax to a stationary state~\cite{pavliotis2014stochastic,hansen2024trek}. Eq.~\eqref{eq:SDEs_methods} admits the analytical solution
\begin{equation}
\bm{X}_{k,t} = \mathrm{e}^{\bm{A}t}\Bigl(\bm{x}_{k,0} + \bm{A}^{-1}\bm{b}\Bigr) - \bm{A}^{-1}\bm{b} 
+\!\!\int_{0}^{t}\mathrm{e}^{\bm{A}(t-\tau)} \bm{B}_{k}\,\mathrm{d}\bm{W}(\tau).
\end{equation}
At long times, the influence of the initial condition decays and the process converges to a stationary complex Gaussian process with mean~\cite{serafini2023quantum}
\begin{equation}
\mathbb{E}[\bm{X}_{k,\infty}] = -\,\bm{A}^{-1}\bm{b}
\end{equation}
and covariance
\begin{equation}
\bm{C}_\infty = \int_{0}^{\infty} \mathrm{e}^{\bm{A}s}\,\bm{B}_{k}\bm{B}_{k}^\dagger\,\mathrm{e}^{\bm{A}^\dagger s}\,\mathrm{d}s\, ,
\end{equation}
which is the unique Hermitian solution of the Lyapunov equation
\begin{equation}
\bm{A}\,\bm{C}_\infty + \bm{C}_\infty \bm{A}^\dagger + \bm{B}_{k}\bm{B}_{k}^\dagger = \bm{0}\, .
\end{equation}
Thus the stationary dynamics are fully described by a complex Gaussian distribution with mean $-\bm{A}^{-1}\bm{b}$ and covariance $\bm{C}_\infty$.

In conclusion, we note that in SDE theory, the diffusion matrix is often defined as $\bm{D} = \bm{B}\bm{B}^{\mathrm{T}}$, such that the Itô isometry yields $\mathbb{E}[\mathrm{d}\bm{X}_{t}\,\mathrm{d}\bm{X}_{t}^{\mathrm{T}}] = \bm{D}\,\mathrm{d}t$~\cite{oksendal2013stochastic}. However, in our setting this relation does not hold individually for each process. Instead, we find that the noise cross-covariance between the two processes is given by
\begin{equation}\label{eq:ito_isometry_methods}
    \mathbb{E}[\mathrm{d}\bm{X}_{t,1}\,\mathrm{d}\bm{X}_{t,2}^{\mathrm{T}}] = \bm{B}_{1}\bm{B}_{2}^{\mathrm{T}}\,\mathrm{d}t=\bm{D}\,\mathrm{d}t\,,
\end{equation}
which defines a shared diffusion structure through the matrix $\bm{D}$. 
\subsection{Numerical simulation methods}
For the numerical solution scheme, we now adapt a notation more common in physics and introduce the index $n$, which denotes the number of walkers, i.e., the number of stochastic trajectories
\begin{equation} \label{eq:SDE_physics_style_methods}
    \partial_t \bm{s}^{(n)}_k(t) = \bm{A} \bm{s}^{(n)}_k(t) + \bm{b} + \bm{B}_{k} \bm{\xi}^{(n)}(t)\,, \quad 
    \begin{aligned}
        k &= 1,2 \,, \\
        n &= 1, \dots, N\,,
    \end{aligned}
\end{equation}
where $\bm{s}_{k}=(s_{k,-}\,\,s_{k,+}\,\,s_{k,z})^{\mathrm{T}}$, and the formal derivative $\bm{\xi}(t) := {\mathrm{d}\bm{W}(t)}/{dt}$ 
is a real-valued vector white noise process in the sense of generalized functions
$\mathbb{E}\bigl[\bm{\xi}(t)\,\bm{\xi}(t')^{\mathrm{T}}\bigr]
= \delta(t - t')\,\mathds{1}_{3}$
and each component of $\bm{\xi}(t)$ has zero mean~\cite{lim2024stochastic}. Comparing Eq.~\eqref{eq:SDE_physics_style_methods} with Eq.~\eqref{eq:bloch_vector_methods} suggests interpreting Eq.~\eqref{eq:SDE_physics_style_methods} as a stochastic Bloch vector. We solve Eq.~\eqref{eq:SDE_physics_style} numerically by an exponential Euler-Maruyama method
\begin{equation}\label{eq:numerical_simulation_stochastic_methods}
    \bm{s}_{k}^{(n)}(m+1) = \mathrm{e}^{\bm{A}\Delta t}\bm{s}_{k}^{(n)}(m) + \bm{b}\Delta t + \sqrt{\Delta t}\,\bm{B}_{k}\,\bm{\xi}^{(n)}(m)\,,
\end{equation}
with $ \bm{\xi}(m) \sim \mathcal{N}(\bm{0}, \mathds{1}_{3})\,, \text{i.i.d. over } m$ and $n=1,...,N$. The deterministic part is treated exactly with the matrix exponential, and the stochastic part is integrated using the standard Euler-Maruyama method~\cite{pavliotis2014stochastic}. We construct the steady state fluctuations by
\begin{equation} \label{eq:stochastic_steady_state_fluctuations_methods}
\mathcal{C}_{ij}^{\mathrm{sto}}(\tau) = \overline{  s_{i,1}^{(n)}(\tau) \,  s_{j,2}^{(n)}(0)  }_{\mathrm{s.s.}} - \overline{  s_{i,1}^{(n)} }_{\mathrm{s.s.}}\, \overline{s_{j,2}^{(n)}}_{\mathrm{s.s.}}\,,
\end{equation}
where $\tau = m\Delta t$ and $\overline{(\cdot)}_{s.s.}$ denotes an average over independent stochastic realizations (walkers) in the steady state. The simulation parameters used for Fig.~\ref{fig:FDTD_plot} are: $\Omega_{\mathrm{R}}=8$, $\Delta=-0.9,0,0.9$, $T_{1}=1$, $T_{2}=2\cdot T_{1}$ and $N=\num[print-unity-mantissa=false]{1e5}$.

\begin{table}
    \caption{Simulation parameters}
    \centering
     \renewcommand{\arraystretch}{2} %
      \resizebox{0.48\textwidth}{!}{ 
    \begin{tabular}{|c||c|c|c|c|c|c|} \hline
    \textbf{Description} & \textbf{Symbol} & \textbf{Fig.~\ref{fig:theoretical_Mollow_spectra_colormap} (a)-(c)} & \textbf{Fig.~\ref{fig:theoretical_Mollow_spectra_colormap}(d)-(f)} & \textbf{Fig.~\ref{fig:theoretical_Mollow_spectra_lineplot}} & \textbf{Fig.~\ref{fig:power_dependent_Mollow_spectrum}} &\textbf{Fig.~\ref{fig:detuning_dependent_Mollow_spectrum}} \\ \hline\hline

    Rabi energy  & $\hbar\Omega_{\mathrm{R}}\, (\si{\micro\electronvolt})$ &$30$  & $ [0, 60 ]\,$ & $30\, $  & (a) $[6.4, 57.7 ]$  &  $21$ 
    
    \\ \hline

    detuning energy   & $\hbar\Delta \,(\si{\micro\electronvolt})$&  $[-60, 60 ]\,$  &  $0$&  $[-10, 10] $     & $0$ &  $[-42, 45]$   \\ \hline

    relaxation time     & $T_{1}\,(\si{\pico\second})$ & 400 &400 & 400  & 450  & 450  \\ \hline

    dephasing time     & $T_{2}\,(\si{\pico\second})$ & 800  & 800 & 800 & 1500  & 900   \\ \hline
    
    number of walkers      & $N$ & \num{4e3} &  \num{4e3} & \num{4e3}, \num[print-unity-mantissa=false]{1e5} & \num{4e3} & \num{15e4} \\ \hline
    \end{tabular} }
    \label{Simulation parameters}
\end{table}

\subsection{Experimental methods}\label{sec:experimental_setup_methods}

Resonance fluorescence measurements were performed under resonant excitation of a single quantum dot. The setup was designed to separate the reflected laser from the emission signal using cross-polarized filtering. The sample was cooled to \SI{4.2}{\kelvin} in a liquid helium bath cryostat.

The sample was mounted on a piezoelectric positioning system, allowing 3D movement with \SI{50}{\nano\meter} precision, and contacted via high-frequency connectors for voltage control of the Schottky diode. Excitation and collection were performed through a high-numerical aperture ($\mathrm{NA}=0.77$) objective.

The optical head consisted of three stages: imaging, excitation, and detection. Excitation light, delivered via an optical fiber, was linearly polarized and reflected by a polarizing beamsplitter to illuminate the sample. Emitted resonance fluorescence photons were collected through the same objective, passed through the polarizing beamsplitter and additional polarizers, and coupled into a single-mode fiber for detection by a Fabry–Perot cavity. More precisely, the QD emission was detected with a scanning Fabry-Perot interferometer and a single photon avalanche diode. A quarter-wave plate above the cryostat window corrected for polarization distortions, and the fiber coupling acted as a spatial filter to further suppress residual laser light.

\section{Acknowledgments}
The authors acknowledge the financial support by Deutsche Forschungsgemeinschaft (DFG) under the DFG DACH project (Project No. 471080402), by the FWF Project No. I 5682: “Cavity-assisted non-classical light generation,” and by the European Space Agency (ESA) Discovery EISI Project No. I-2023-02655: “Simulation toolbox for unconditionally secure on-chip satellite quantum communication networks operating in the telecom wavelength range."  The authors acknowledge the financial support by the University of Graz.  F.H. and U.H. thank Norbert Mauser for helpful discussions. F.H. thanks Andrew Lingenfelter for valuable discussions.
\subsection{Data availability}
Data underlying the results presented in this paper are not publicly
available at this time but may be obtained from the authors upon reasonable request.

\subsection{Competing interests}
The authors declare no competing interests.

\bibliographystyle{apsrev4-2}
\bibliography{references}

\end{document}